\documentclass[twoside,twocolumn,english,prl,superscriptaddress,nobibnotes,nofootinbib]{revtex4-1}
\pdfoutput=1
\usepackage[latin9]{inputenc}
\usepackage{color}
\usepackage{babel}
\usepackage{verbatim}
\usepackage{amsmath}
\usepackage{amssymb}
\usepackage[unicode=true,
 bookmarks=true,bookmarksnumbered=true,bookmarksopen=false,
 breaklinks=false,pdfborder={0 0 0},backref=false,colorlinks=true]
 {hyperref}
\hypersetup{pdftitle={A Synopsis of the Minimal Modal Interpretation of Quantum Theory},
 pdfauthor={Jacob A. Barandes, David Kagan},
 pdfsubject={Physics},
 pdfkeywords={physics, quantum mechanics, interpretations}}

\makeatletter

\@ifundefined{textcolor}{}
{%
 \definecolor{BLACK}{gray}{0}
 \definecolor{WHITE}{gray}{1}
 \definecolor{RED}{rgb}{1,0,0}
 \definecolor{GREEN}{rgb}{0,1,0}
 \definecolor{BLUE}{rgb}{0,0,1}
 \definecolor{CYAN}{cmyk}{1,0,0,0}
 \definecolor{MAGENTA}{cmyk}{0,1,0,0}
 \definecolor{YELLOW}{cmyk}{0,0,1,0}
}

\usepackage{url}
\usepackage{graphicx}
\usepackage{tikz}
\usetikzlibrary{matrix,arrows,decorations.pathmorphing,decorations.pathreplacing}

\makeatother

\begin{document}

\title{A Synopsis of the Minimal Modal Interpretation of Quantum Theory}

\author{Jacob A. Barandes}

\email{barandes@physics.harvard.edu}

\affiliation{Jefferson Physical Laboratory, Harvard University, Cambridge, MA
02138}

\author{David Kagan}

\email{dkagan@umassd.edu}

\affiliation{Department of Physics, University of Massachusetts Dartmouth, North
Dartmouth, MA 02747}

\date{\today}
\begin{abstract}
We summarize a new realist interpretation of quantum theory that builds
on the existing physical structure of the theory and allows experiments
to have definite outcomes, but leaves the theory's basic dynamical
content essentially intact. Much as classical systems have specific
states that evolve along definite trajectories through configuration
spaces, the traditional formulation of quantum theory asserts that
closed quantum systems have specific states that evolve unitarily
along definite trajectories through Hilbert spaces, and our interpretation
extends this intuitive picture of states and Hilbert-space trajectories
to the case of open quantum systems as well. Our interpretation---which
we claim is ultimately compatible with Lorentz invariance---reformulates
wave-function collapse in terms of an underlying interpolating dynamics,
makes it possible to derive the Born rule from deeper principles,
and resolves several open questions regarding ontological stability
and dynamics.
\end{abstract}
\maketitle

\noindent \begin{center}

\global\long\def\hyphen{\mbox{-}}
\global\long\def\colon{\mbox{:}}

\global\long\def\difficult{\dagger}
\global\long\def\anc{*}

\global\long\def\exampleclose{\blacklozenge}
\global\long\def\solutionclose{\blacksquare}
\global\long\def\stepclose{\checkmark}

\global\long\def\fakespace{}

\global\long\def\subarrayleftleft#1#2{\begin{subarray}{l}
 #1\\
#2 
\end{subarray}}

\global\long\def\quote#1{``#1"}

\global\long\def\eqsbrace#1{\left.#1\qquad\qquad\qquad\right\}  }

\global\long\def\presup#1#2{\vphantom{#1}^{#2}#1}
\global\long\def\presub#1#2{\vphantom{#1}_{#2}#1}

\global\long\def\tight#1{\!\negthickspace\negthickspace#1\negthickspace\negthickspace\!}
\global\long\def\const{\mathrm{const}}
\global\long\def\degrees{\circ}

\global\long\def\units#1{\mathrm{#1}}
\global\long\def\hasunits#1{\left[#1\right]}

\global\long\def\from{\leftarrow}
\global\long\def\xto#1#2{\xrightarrow[#1]{#2}}
\global\long\def\xfrom#1#2{\xleftarrow[#1]{#2}}
\global\long\def\mapsfrom{\from}
\global\long\def\xmapsto#1#2{\overset{{\scriptstyle #2}}{\underset{{\scriptstyle #1}}{\mapsto}}}

\global\long\def\impliedby{\Longleftarrow}
\global\long\def\exchange{\leftrightarrow}

\global\long\def\binaryand{\mathrm{\ and\ }}
\global\long\def\binaryor{\mathrm{\ or\ }}
\global\long\def\binaryis{\mathrm{\ is\ }}
\global\long\def\binaryfor{\mathrm{\ for\ }}
\global\long\def\binaryfrom{\mathrm{\ from\ }}
\global\long\def\binarytext#1{\mathrm{\ #1\ }}

\global\long\def\whichfor{\mathrm{for\ }}
\global\long\def\whichif{\mathrm{if\ }}
\global\long\def\whichotherwise{\mathrm{otherwise}}
\global\long\def\whichtext#1{\mathrm{#1\ }}
\global\long\def\also{\mathrm{also}}

\global\long\def\given{\vert}

\global\long\def\QED{\mathrm{QED}}

\global\long\def\mapping#1#2#3{#1:#2\to#3}
\global\long\def\composition{\circ}

\global\long\def\set#1{\left\{  #1\right\}  }
\global\long\def\setindexed#1#2{\left\{  #1\right\}  _{#2}}

\global\long\def\setbuild#1#2{\left\{  \left.\!#1\right|#2\right\}  }
\global\long\def\complem{\mathrm{c}}

\global\long\def\union{\cup}
\global\long\def\intersection{\cap}
\global\long\def\cartesianprod{\times}
\global\long\def\disjointunion{\sqcup}

\global\long\def\isomorphic{\cong}

\global\long\def\sgn{\mathrm{sgn\,}}
\global\long\def\absval#1{\left|#1\right|}

\global\long\def\average#1{\left\langle #1\right\rangle }
\global\long\def\saverage#1{\left\langle #1\right\rangle }
\global\long\def\taverage#1{\overline{#1}}
\global\long\def\mean#1{\overline{#1}}

\global\long\def\evaluate#1{\left.#1\right|}

\global\long\def\notpropto{\not\propto}
\global\long\def\floor#1{\left\lfloor #1\right\rfloor }
\global\long\def\ceil#1{\left\lceil #1\right\rceil }

\global\long\def\integration#1#2{#1#2\;}
\global\long\def\differential#1{#1\;}

\global\long\def\csch{\mathrm{csch\,}}
\global\long\def\arccot{\mathrm{arccot\,}}
\global\long\def\oforder#1{\mathcal{O}\left(#1\right)}

\global\long\def\convolution{\circ}

\global\long\def\conj{\ast}
\global\long\def\re{\mathrm{Re\,}}
\global\long\def\im{\mathrm{Im\,}}

\global\long\def\Disc{\mathrm{Disc\,}}
\global\long\def\Res{\mathrm{Res\,}}

\global\long\def\Span{\mathrm{Span\,}}
\global\long\def\Tr{\mathrm{Tr\,}}
\global\long\def\tr{\mathrm{tr\,}}

\global\long\def\transp{\mathrm{T}}
\global\long\def\adj{\dagger}

\global\long\def\hasrep{\rightarrowtail}
\global\long\def\repfor{\leftarrowtail}

\global\long\def\xhasrep#1#2{\overset{#2}{\underset{#1}{\rightarrowtail}}}
\global\long\def\xrepfor#1#2{\overset{#2}{\underset{#1}{\leftarrowtail}}}

\global\long\def\diag#1{\mathrm{diag}\left(#1\right)}

\global\long\def\directsum{\oplus}
\global\long\def\directprod{\otimes}
\global\long\def\tensorprod{\otimes}

\global\long\def\boundary{\partial}

\global\long\def\notparallel{\not\parallel}
\global\long\def\notperp{\not\perp}

\global\long\def\svec#1{\vec{#1}}
\global\long\def\norm#1{\absval{#1}}
\global\long\def\stensor#1{\overleftrightarrow{#1}}
\global\long\def\dotprod{\cdot}
\global\long\def\crossprod{\times}

\global\long\def\del{\svec{\nabla}}
\global\long\def\dalemb{\square}

\global\long\def\gradient{\svec{\nabla}}
\global\long\def\divergence{\svec{\nabla}\cdot}
\global\long\def\curl{\svec{\nabla}\times}
\global\long\def\laplacian{\svec{\nabla}^{2}}

\global\long\def\gradientprime{\svec{\nabla}^{\prime}}
\global\long\def\divergenceprime{\svec{\nabla}^{\prime}\cdot}
\global\long\def\curlprime{\svec{\nabla}^{\prime}\times}
\global\long\def\laplacianprime{\svec{\nabla}^{\prime2}}

\global\long\def\homeomorphic{\cong}
\global\long\def\diffeomorphic{\cong}

\global\long\def\dual{\ast}
\global\long\def\extderiv{\mathrm{d}}
\global\long\def\hodge{\mathrm{\star}}

\global\long\def\hodgeprod#1#2{\left(#1,#2\right)_{\star}}

\global\long\def\liederiv#1#2{\pounds_{#2}#1}

\global\long\def\liebracket#1#2{\left[#1,#2\right]}

\global\long\def\connection{P}
\global\long\def\cov{\nabla}

\global\long\def\tud#1#2#3{#1_{\phantom{#2}#3}^{#2}}

\global\long\def\tdu#1#2#3{#1_{#2}^{\phantom{#2}#3}}

\global\long\def\tudu#1#2#3#4{#1_{\phantom{#2}#3}^{#2\phantom{#3}#4}}

\global\long\def\tdud#1#2#3#4{#1_{#2\phantom{#3}#4}^{\phantom{#2}#3}}

\global\long\def\tudud#1#2#3#4#5{#1_{\phantom{#2}#3\phantom{#4}#5}^{#2\phantom{#3}#4}}

\global\long\def\tdudu#1#2#3#4#5{#1_{#2\phantom{#3}#4}^{\phantom{#2}#3\phantom{#4}#5}}

\global\long\def\Probability#1{\mathrm{Prob}\left(#1\right)}
\global\long\def\Amplitude#1{\mathrm{Amp}\left(#1\right)}

\global\long\def\expectval#1{\left\langle #1\right\rangle }
\global\long\def\op#1{\hat{#1}}

\global\long\def\bra#1{\left\langle #1\right|}
\global\long\def\ket#1{\left|#1\right\rangle }
\global\long\def\braket#1#2{\left\langle \left.\!#1\right|#2\right\rangle }

\global\long\def\dbbra#1{\left\langle #1\right\Vert }
\global\long\def\dbket#1{\left\Vert #1\right\rangle }

\global\long\def\inprod#1#2{\left\langle #1,#2\right\rangle }
\global\long\def\normket#1{\left\Vert #1\right\Vert }

\global\long\def\comm#1#2{\left[#1,#2\right]}
\global\long\def\acomm#1#2{\left\{  #1,#2\right\}  }

\global\long\def\pbrack#1#2{\left\{  #1,#2\right\}  _{\mathrm{PB}}}

\global\long\def\timeorder{\mathrm{T}}

\par\end{center}

\section{Introduction\label{sec:Introduction}}

In this letter and in a more comprehensive companion paper \cite{BarandesKagan:2014mmiqt},
we present a realist interpretation of quantum theory that hews closely
to the basic structure of the theory in its widely accepted current
form. Our primary goal is to move beyond instrumentalism and describe
an actual reality that lies behind the mathematical formalism of quantum
theory. We also intend to provide new hope to those who find themselves
disappointed with the pace of progress on making sense of the theory's
foundations \cite{Weinberg:2013qasw,Weinberg:2012loqm}.

\subsection{Why Do We Need a New Interpretation?\label{sub:Why-Do-We-Need-a-New-Interpretation}}

\begin{figure}
\begin{centering}
\begin{center}
\scalebox{0.8}{
\definecolor{cffffff}{RGB}{255,255,255}

\begin{tikzpicture}[y=0.60pt,x=0.60pt,yscale=-1, inner sep=0pt, outer sep=0pt]
\begin{scope}[shift={(-47.98725,-39.68933)}]
  \begin{scope}[cm={{0.53212,0.0,0.0,0.53212,(-88.27664,-60.44546)}}]
    \path[shift={(-12.5874,170.91883)},draw=black,miter limit=4.00,line
      width=0.800pt] (400.0000,237.3622) .. controls (400.0000,250.3803) and
      (389.4467,260.9336) .. (376.4286,260.9336) .. controls (363.4104,260.9336) and
      (352.8571,250.3803) .. (352.8571,237.3622) .. controls (352.8571,224.3440) and
      (363.4104,213.7908) .. (376.4286,213.7908) .. controls (389.4467,213.7908) and
      (400.0000,224.3440) .. (400.0000,237.3622) -- cycle;
    \path[draw=black,line join=miter,line cap=butt,line width=0.800pt]
      (351.0280,388.1869) .. controls (357.1764,401.3189) and (358.3277,414.4508) ..
      (351.0280,427.5828);
    \path[draw=black,line join=miter,line cap=butt,line width=0.800pt]
      (376.2826,388.6920) .. controls (370.1342,401.8239) and (368.9829,414.9559) ..
      (376.2826,428.0879);
    \path[draw=black,line join=miter,line cap=butt,line width=0.800pt]
      (350.5229,393.7427) -- (355.5737,390.4597);
    \path[draw=black,line join=miter,line cap=butt,line width=0.800pt]
      (351.7856,398.5410) -- (357.0889,395.2579);
    \path[draw=black,line join=miter,line cap=butt,line width=0.800pt]
      (352.5433,402.0765) -- (358.3516,401.5714);
    \path[draw=black,line join=miter,line cap=butt,line width=0.800pt]
      (353.5534,406.6222) -- (358.3516,407.3798);
    \path[draw=black,line join=miter,line cap=butt,line width=0.800pt]
      (353.0483,411.1678) -- (357.8465,413.1882);
    \path[draw=black,line join=miter,line cap=butt,line width=0.800pt]
      (352.2907,416.7237) -- (357.0889,417.9864);
    \path[draw=black,line join=miter,line cap=butt,line width=0.800pt]
      (350.7755,420.7643) -- (356.0788,422.7846);
    \path[draw=black,line join=miter,line cap=butt,line width=0.800pt]
      (349.5128,424.2998) -- (354.8161,427.0778);
    \path[draw=black,line join=miter,line cap=butt,line width=0.800pt]
      (372.9988,390.4597) -- (377.5445,393.4902);
    \path[draw=black,line join=miter,line cap=butt,line width=0.800pt]
      (371.7362,395.0054) -- (376.2818,396.7732);
    \path[draw=black,line join=miter,line cap=butt,line width=0.800pt]
      (370.2209,398.5410) -- (375.0191,400.3087);
    \path[draw=black,line join=miter,line cap=butt,line width=0.800pt]
      (369.2108,401.8239) -- (373.5039,403.5917);
    \path[draw=black,line join=miter,line cap=butt,line width=0.800pt]
      (368.7057,407.1272) -- (373.7565,407.1272);
    \path[draw=black,line join=miter,line cap=butt,line width=0.800pt]
      (369.2108,411.1678) -- (374.5141,410.4102);
    \path[draw=black,line join=miter,line cap=butt,line width=0.800pt]
      (368.9582,416.9762) -- (374.5141,415.9661);
    \path[draw=black,line join=miter,line cap=butt,line width=0.800pt]
      (370.4735,420.7643) -- (375.0191,419.5016);
    \path[draw=black,line join=miter,line cap=butt,line width=0.800pt]
      (372.4938,424.2998) -- (376.0293,422.2795);
    \path[draw=black,line join=miter,line cap=butt,line width=0.800pt]
      (373.5039,427.3303) -- (377.2920,424.8049);
    \path[draw=black,line join=miter,line cap=butt,line width=0.800pt]
      (321.2285,400.8138) .. controls (287.3022,407.3572) and (269.0644,425.6670) ..
      (256.5788,448.2910);
    \path[draw=black,line join=miter,line cap=butt,line width=0.800pt]
      (336.3808,424.0473) .. controls (307.7466,431.3340) and (292.4566,447.5169) ..
      (279.8123,465.4635);
    \path[draw=black,line join=miter,line cap=butt,line width=0.800pt]
      (357.5940,444.2504) .. controls (332.8225,452.8230) and (317.4292,467.0225) ..
      (306.0762,483.6463);
  \end{scope}
  \begin{scope}[cm={{0.53212,0.0,0.0,0.53212,(82.36707,6.48496)}}]
    \path[draw=black,line join=miter,line cap=butt,line width=0.800pt]
      (366.6854,201.8137) -- (366.6854,170.4990);
    \path[draw=black,line join=miter,line cap=butt,line width=0.800pt]
      (394.9696,207.8747) -- (413.1524,181.6107);
    \path[draw=black,line join=miter,line cap=butt,line width=0.800pt]
      (408.1016,230.0980) -- (435.3757,214.9457);
    \path[draw=black,line join=miter,line cap=butt,line width=0.800pt]
      (410.1219,249.2909) -- (437.3961,248.2807);
    \path[draw=black,line join=miter,line cap=butt,line width=0.800pt]
      (401.0306,277.5752) -- (420.2235,289.6970);
    \path[draw=black,line join=miter,line cap=butt,line width=0.800pt]
      (386.8275,290.7072) -- (395.9189,311.9204);
    \path[draw=black,line join=miter,line cap=butt,line width=0.800pt]
      (366.6448,293.7376) -- (366.6448,321.0117);
    \path[draw=black,line join=miter,line cap=butt,line width=0.800pt]
      (338.4011,292.7275) -- (321.2285,308.8899);
    \path[draw=black,line join=miter,line cap=butt,line width=0.800pt]
      (323.2488,271.5143) -- (302.0356,278.5853);
    \path[draw=black,line join=miter,line cap=butt,line width=0.800pt]
      (317.1879,242.2198) -- (288.9036,231.1082);
    \path[draw=black,line join=miter,line cap=butt,line width=0.800pt]
      (327.2894,213.9356) -- (307.0864,195.7528);
    \path[draw=black,line join=miter,line cap=butt,line width=0.800pt]
      (346.4823,208.8848) -- (329.3097,180.6005);
    \path[shift={(111.11678,102.02541)},draw=black,miter limit=4.00,line
      width=0.800pt] (426.2844,163.4280) .. controls (426.2844,196.3436) and
      (349.8523,223.0269) .. (255.5686,223.0269) .. controls (161.2849,223.0269) and
      (84.8528,196.3436) .. (84.8528,163.4280) .. controls (84.8528,130.5123) and
      (161.2849,103.8290) .. (255.5686,103.8290) .. controls (349.8523,103.8290) and
      (426.2844,130.5123) .. (426.2844,163.4280) -- cycle;
    \path[shift={(104.04571,77.78175)},draw=black,miter limit=4.00,line
      width=0.800pt] (458.6093,190.7021) .. controls (458.6093,230.3124) and
      (371.0968,262.4229) .. (263.1447,262.4229) .. controls (155.1927,262.4229) and
      (67.6802,230.3124) .. (67.6802,190.7021) .. controls (67.6802,151.0917) and
      (155.1927,118.9812) .. (263.1447,118.9812) .. controls (371.0968,118.9812) and
      (458.6093,151.0917) .. (458.6093,190.7021) -- cycle;
    \path[shift={(104.04571,77.78175)},draw=black,miter limit=4.00,line
      width=0.800pt] (483.8631,193.2275) .. controls (483.8631,237.5799) and
      (385.7225,273.5346) .. (264.6600,273.5346) .. controls (143.5974,273.5346) and
      (45.4569,237.5799) .. (45.4569,193.2275) .. controls (45.4569,148.8750) and
      (143.5974,112.9203) .. (264.6600,112.9203) .. controls (385.7225,112.9203) and
      (483.8631,148.8750) .. (483.8631,193.2275) -- cycle;
    \path[shift={(104.04571,77.78175)},draw=black,miter limit=4.00,line
      width=0.800pt] (342.4417,183.6310) .. controls (342.4417,200.3678) and
      (304.9041,213.9356) .. (258.5991,213.9356) .. controls (212.2940,213.9356) and
      (174.7564,200.3677) .. (174.7564,183.6310) .. controls (174.7564,166.8942) and
      (212.2940,153.3264) .. (258.5991,153.3264) .. controls (304.9041,153.3264) and
      (342.4417,166.8942) .. (342.4417,183.6310) -- cycle;
    \path[shift={(97.98479,75.76144)},draw=black,miter limit=4.00,line
      width=0.800pt] (366.6854,186.6615) .. controls (366.6854,207.3034) and
      (320.7809,224.0371) .. (264.1549,224.0371) .. controls (207.5288,224.0371) and
      (161.6244,207.3034) .. (161.6244,186.6615) .. controls (161.6244,166.0195) and
      (207.5289,149.2858) .. (264.1549,149.2858) .. controls (320.7809,149.2858) and
      (366.6854,166.0195) .. (366.6854,186.6615) -- cycle;
    \path[shift={(201.02036,92.93403)},draw=black,fill=cffffff,miter limit=4.00,line
      width=0.800pt] (201.0204,155.3467) .. controls (201.0204,175.4308) and
      (184.7390,191.7122) .. (164.6549,191.7122) .. controls (144.5708,191.7122) and
      (128.2894,175.4308) .. (128.2894,155.3467) .. controls (128.2894,135.2626) and
      (144.5708,118.9812) .. (164.6549,118.9812) .. controls (184.7390,118.9812) and
      (201.0204,135.2626) .. (201.0204,155.3467) -- cycle;
    \path[shift={(110.10663,81.82236)},draw=black,fill=cffffff,miter limit=4.00,line
      width=0.800pt] (434.3656,185.6513) .. controls (434.3656,190.6723) and
      (430.2953,194.7427) .. (425.2742,194.7427) .. controls (420.2532,194.7427) and
      (416.1829,190.6723) .. (416.1829,185.6513) .. controls (416.1829,180.6303) and
      (420.2532,176.5599) .. (425.2742,176.5599) .. controls (430.2953,176.5599) and
      (434.3656,180.6303) .. (434.3656,185.6513) -- cycle;
    \path[shift={(104.04571,77.78175)},draw=black,fill=cffffff,miter limit=4.00,line
      width=0.800pt] (355.5737,255.8569) .. controls (355.5737,259.4832) and
      (352.6340,262.4229) .. (349.0077,262.4229) .. controls (345.3814,262.4229) and
      (342.4417,259.4832) .. (342.4417,255.8569) .. controls (342.4417,252.2306) and
      (345.3814,249.2909) .. (349.0077,249.2909) .. controls (352.6340,249.2909) and
      (355.5737,252.2306) .. (355.5737,255.8569) -- cycle;
    \path[shift={(104.04571,77.78175)},draw=black,fill=cffffff,miter limit=4.00,line
      width=0.800pt] (57.5787,210.4000) .. controls (57.5787,214.5842) and
      (54.1867,217.9762) .. (50.0026,217.9762) .. controls (45.8184,217.9762) and
      (42.4264,214.5842) .. (42.4264,210.4000) .. controls (42.4264,206.2158) and
      (45.8184,202.8239) .. (50.0026,202.8239) .. controls (54.1867,202.8239) and
      (57.5787,206.2158) .. (57.5787,210.4000) -- cycle;
    \path[shift={(106.06602,76.77159)},draw=black,fill=cffffff,miter limit=4.00,line
      width=0.800pt] (159.6041,193.7325) .. controls (159.6041,196.5220) and
      (157.3428,198.7833) .. (154.5533,198.7833) .. controls (151.7639,198.7833) and
      (149.5026,196.5220) .. (149.5026,193.7325) .. controls (149.5026,190.9430) and
      (151.7639,188.6817) .. (154.5533,188.6817) .. controls (157.3428,188.6817) and
      (159.6041,190.9430) .. (159.6041,193.7325) -- cycle;
    \path[shift={(104.04571,77.78175)},draw=black,fill=cffffff,miter limit=4.00,line
      width=0.800pt] (181.8275,187.6716) .. controls (181.8275,190.4611) and
      (179.5662,192.7224) .. (176.7767,192.7224) .. controls (173.9872,192.7224) and
      (171.7259,190.4611) .. (171.7259,187.6716) .. controls (171.7259,184.8821) and
      (173.9872,182.6208) .. (176.7767,182.6208) .. controls (179.5662,182.6208) and
      (181.8275,184.8821) .. (181.8275,187.6716) -- cycle;
  \end{scope}
  \begin{scope}[cm={{0.53212,0.0,0.0,0.53212,(-72.15105,33.08298)}}]
    \path[cm={{1.00135,0.0,0.0,0.52346,(45.00199,57.46502)}},draw=black,line
      join=miter,line cap=butt,line width=0.800pt] (327.2894,49.2807) .. controls
      (332.8606,49.1288) and (330.2429,56.7156) .. (327.5420,58.5404) .. controls
      (320.2226,63.4855) and (311.2492,56.8977) .. (308.7700,49.7858) .. controls
      (304.3351,37.0642) and (314.3658,24.4320) .. (326.5318,21.5015) .. controls
      (344.3859,17.2009) and (361.1068,31.0650) .. (364.3284,48.2706) .. controls
      (368.6222,71.2029) and (350.7973,92.1283) .. (328.5521,95.5794) .. controls
      (300.5587,99.9221) and (275.3809,78.0820) .. (271.7310,50.7959) .. controls
      (267.3106,17.7487) and (293.1937,-11.7062) .. (325.5217,-15.5374) .. controls
      (363.6191,-20.0524) and (397.3656,9.8899) .. (401.3673,47.2604) .. controls
      (405.9874,90.4062) and (371.9757,128.4534) .. (329.5622,132.6183) .. controls
      (281.3694,137.3507) and (239.0151,99.2627) .. (234.6921,51.8061) .. controls
      (229.8424,-1.4332) and (272.0115,-48.0991) .. (324.5115,-52.5763) .. controls
      (382.7966,-57.5470) and (433.7774,-11.2934) .. (438.4062,46.2503);
    \path[draw=black,line join=miter,line cap=butt,line width=0.800pt]
      (470.7311,73.5244) -- (462.6499,112.9203);
    \path[draw=black,line join=miter,line cap=butt,line width=0.800pt]
      (476.7920,35.1386) -- (499.0154,71.5041);
    \path[draw=black,line join=miter,line cap=butt,line width=0.800pt]
      (371.7361,19.9863) -- (409.1118,12.9152);
    \path[draw=black,line join=miter,line cap=butt,line width=0.800pt]
      (267.6904,41.1995) -- (295.9747,13.9254);
    \path[draw=black,line join=miter,line cap=butt,line width=0.800pt]
      (261.6295,102.8188) -- (288.9036,122.0117);
    \path[draw=black,line join=miter,line cap=butt,line width=0.800pt]
      (360.6245,134.1335) -- (413.1524,141.2046);
    \path[draw=black,line join=miter,line cap=butt,line width=0.800pt]
      (381.8377,91.7071) -- (401.0306,72.5142);
    \path[draw=black,line join=miter,line cap=butt,line width=0.800pt]
      (342.4417,62.4127) -- (320.2184,81.6056);
    \path[draw=black,fill=cffffff,miter limit=4.00,line width=0.800pt]
      (399.0103,45.7452) .. controls (399.0103,47.6978) and (397.4273,49.2807) ..
      (395.4747,49.2807) .. controls (393.5221,49.2807) and (391.9392,47.6978) ..
      (391.9392,45.7452) .. controls (391.9392,43.7925) and (393.5221,42.2096) ..
      (395.4747,42.2096) .. controls (397.4273,42.2096) and (399.0103,43.7925) ..
      (399.0103,45.7452) -- cycle;
    \path[shift={(37.88071,8.5863)},draw=black,fill=cffffff,miter limit=4.00,line
      width=0.800pt] (399.0103,45.7452) .. controls (399.0103,47.6978) and
      (397.4273,49.2807) .. (395.4747,49.2807) .. controls (393.5221,49.2807) and
      (391.9392,47.6978) .. (391.9392,45.7452) .. controls (391.9392,43.7925) and
      (393.5221,42.2096) .. (395.4747,42.2096) .. controls (397.4273,42.2096) and
      (399.0103,43.7925) .. (399.0103,45.7452) -- cycle;
    \path[shift={(23.73858,18.68782)},draw=black,fill=cffffff,miter limit=4.00,line
      width=0.800pt] (399.0103,45.7452) .. controls (399.0103,47.6978) and
      (397.4273,49.2807) .. (395.4747,49.2807) .. controls (393.5221,49.2807) and
      (391.9392,47.6978) .. (391.9392,45.7452) .. controls (391.9392,43.7925) and
      (393.5221,42.2096) .. (395.4747,42.2096) .. controls (397.4273,42.2096) and
      (399.0103,43.7925) .. (399.0103,45.7452) -- cycle;
    \path[shift={(-9.59646,32.82996)},draw=black,fill=cffffff,miter limit=4.00,line
      width=0.800pt] (399.0103,45.7452) .. controls (399.0103,47.6978) and
      (397.4273,49.2807) .. (395.4747,49.2807) .. controls (393.5221,49.2807) and
      (391.9392,47.6978) .. (391.9392,45.7452) .. controls (391.9392,43.7925) and
      (393.5221,42.2096) .. (395.4747,42.2096) .. controls (397.4273,42.2096) and
      (399.0103,43.7925) .. (399.0103,45.7452) -- cycle;
    \path[shift={(-37.88073,7.57614)},draw=black,fill=cffffff,miter limit=4.00,line
      width=0.800pt] (399.0103,45.7452) .. controls (399.0103,47.6978) and
      (397.4273,49.2807) .. (395.4747,49.2807) .. controls (393.5221,49.2807) and
      (391.9392,47.6978) .. (391.9392,45.7452) .. controls (391.9392,43.7925) and
      (393.5221,42.2096) .. (395.4747,42.2096) .. controls (397.4273,42.2096) and
      (399.0103,43.7925) .. (399.0103,45.7452) -- cycle;
    \path[shift={(-64.1447,34.85026)},draw=black,fill=cffffff,miter limit=4.00,line
      width=0.800pt] (399.0103,45.7452) .. controls (399.0103,47.6978) and
      (397.4273,49.2807) .. (395.4747,49.2807) .. controls (393.5221,49.2807) and
      (391.9392,47.6978) .. (391.9392,45.7452) .. controls (391.9392,43.7925) and
      (393.5221,42.2096) .. (395.4747,42.2096) .. controls (397.4273,42.2096) and
      (399.0103,43.7925) .. (399.0103,45.7452) -- cycle;
    \path[shift={(-24.74875,56.06347)},draw=black,fill=cffffff,miter limit=4.00,line
      width=0.800pt] (399.0103,45.7452) .. controls (399.0103,47.6978) and
      (397.4273,49.2807) .. (395.4747,49.2807) .. controls (393.5221,49.2807) and
      (391.9392,47.6978) .. (391.9392,45.7452) .. controls (391.9392,43.7925) and
      (393.5221,42.2096) .. (395.4747,42.2096) .. controls (397.4273,42.2096) and
      (399.0103,43.7925) .. (399.0103,45.7452) -- cycle;
    \path[shift={(27.77918,50.00255)},draw=black,fill=cffffff,miter limit=4.00,line
      width=0.800pt] (399.0103,45.7452) .. controls (399.0103,47.6978) and
      (397.4273,49.2807) .. (395.4747,49.2807) .. controls (393.5221,49.2807) and
      (391.9392,47.6978) .. (391.9392,45.7452) .. controls (391.9392,43.7925) and
      (393.5221,42.2096) .. (395.4747,42.2096) .. controls (397.4273,42.2096) and
      (399.0103,43.7925) .. (399.0103,45.7452) -- cycle;
    \path[shift={(46.97208,37.88072)},draw=black,fill=cffffff,miter limit=4.00,line
      width=0.800pt] (399.0103,45.7452) .. controls (399.0103,47.6978) and
      (397.4273,49.2807) .. (395.4747,49.2807) .. controls (393.5221,49.2807) and
      (391.9392,47.6978) .. (391.9392,45.7452) .. controls (391.9392,43.7925) and
      (393.5221,42.2096) .. (395.4747,42.2096) .. controls (397.4273,42.2096) and
      (399.0103,43.7925) .. (399.0103,45.7452) -- cycle;
    \path[shift={(-43.94165,65.15484)},draw=black,fill=cffffff,miter limit=4.00,line
      width=0.800pt] (399.0103,45.7452) .. controls (399.0103,47.6978) and
      (397.4273,49.2807) .. (395.4747,49.2807) .. controls (393.5221,49.2807) and
      (391.9392,47.6978) .. (391.9392,45.7452) .. controls (391.9392,43.7925) and
      (393.5221,42.2096) .. (395.4747,42.2096) .. controls (397.4273,42.2096) and
      (399.0103,43.7925) .. (399.0103,45.7452) -- cycle;
    \path[shift={(-90.40866,33.84011)},draw=black,fill=cffffff,miter limit=4.00,line
      width=0.800pt] (399.0103,45.7452) .. controls (399.0103,47.6978) and
      (397.4273,49.2807) .. (395.4747,49.2807) .. controls (393.5221,49.2807) and
      (391.9392,47.6978) .. (391.9392,45.7452) .. controls (391.9392,43.7925) and
      (393.5221,42.2096) .. (395.4747,42.2096) .. controls (397.4273,42.2096) and
      (399.0103,43.7925) .. (399.0103,45.7452) -- cycle;
    \path[shift={(-62.12439,-2.52538)},draw=black,fill=cffffff,miter limit=4.00,line
      width=0.800pt] (399.0103,45.7452) .. controls (399.0103,47.6978) and
      (397.4273,49.2807) .. (395.4747,49.2807) .. controls (393.5221,49.2807) and
      (391.9392,47.6978) .. (391.9392,45.7452) .. controls (391.9392,43.7925) and
      (393.5221,42.2096) .. (395.4747,42.2096) .. controls (397.4273,42.2096) and
      (399.0103,43.7925) .. (399.0103,45.7452) -- cycle;
    \path[shift={(-17.67768,-1.51523)},draw=black,fill=cffffff,miter limit=4.00,line
      width=0.800pt] (399.0103,45.7452) .. controls (399.0103,47.6978) and
      (397.4273,49.2807) .. (395.4747,49.2807) .. controls (393.5221,49.2807) and
      (391.9392,47.6978) .. (391.9392,45.7452) .. controls (391.9392,43.7925) and
      (393.5221,42.2096) .. (395.4747,42.2096) .. controls (397.4273,42.2096) and
      (399.0103,43.7925) .. (399.0103,45.7452) -- cycle;
    \path[shift={(-2.02031,-4.04061)},draw=black,fill=cffffff,miter limit=4.00,line
      width=0.800pt] (400.0204,85.6462) .. controls (400.0204,93.4567) and
      (389.8445,99.7883) .. (377.2920,99.7883) .. controls (364.7394,99.7883) and
      (354.5635,93.4567) .. (354.5635,85.6462) .. controls (354.5635,77.8357) and
      (364.7394,71.5041) .. (377.2920,71.5041) .. controls (389.8445,71.5041) and
      (400.0204,77.8357) .. (400.0204,85.6462) -- cycle;
    \path[shift={(-53.03302,48.9924)},draw=black,fill=cffffff,miter limit=4.00,line
      width=0.800pt] (399.0103,45.7452) .. controls (399.0103,47.6978) and
      (397.4273,49.2807) .. (395.4747,49.2807) .. controls (393.5221,49.2807) and
      (391.9392,47.6978) .. (391.9392,45.7452) .. controls (391.9392,43.7925) and
      (393.5221,42.2096) .. (395.4747,42.2096) .. controls (397.4273,42.2096) and
      (399.0103,43.7925) .. (399.0103,45.7452) -- cycle;
    \path[shift={(-86.36805,59.09392)},draw=black,fill=cffffff,miter limit=4.00,line
      width=0.800pt] (399.0103,45.7452) .. controls (399.0103,47.6978) and
      (397.4273,49.2807) .. (395.4747,49.2807) .. controls (393.5221,49.2807) and
      (391.9392,47.6978) .. (391.9392,45.7452) .. controls (391.9392,43.7925) and
      (393.5221,42.2096) .. (395.4747,42.2096) .. controls (397.4273,42.2096) and
      (399.0103,43.7925) .. (399.0103,45.7452) -- cycle;
    \path[shift={(19.69796,76.26652)},draw=black,fill=cffffff,miter limit=4.00,line
      width=0.800pt] (399.0103,45.7452) .. controls (399.0103,47.6978) and
      (397.4273,49.2807) .. (395.4747,49.2807) .. controls (393.5221,49.2807) and
      (391.9392,47.6978) .. (391.9392,45.7452) .. controls (391.9392,43.7925) and
      (393.5221,42.2096) .. (395.4747,42.2096) .. controls (397.4273,42.2096) and
      (399.0103,43.7925) .. (399.0103,45.7452) -- cycle;
    \path[shift={(62.12437,76.26652)},draw=black,fill=cffffff,miter limit=4.00,line
      width=0.800pt] (399.0103,45.7452) .. controls (399.0103,47.6978) and
      (397.4273,49.2807) .. (395.4747,49.2807) .. controls (393.5221,49.2807) and
      (391.9392,47.6978) .. (391.9392,45.7452) .. controls (391.9392,43.7925) and
      (393.5221,42.2096) .. (395.4747,42.2096) .. controls (397.4273,42.2096) and
      (399.0103,43.7925) .. (399.0103,45.7452) -- cycle;
    \path[shift={(72.2259,-24.74874)},draw=black,fill=cffffff,miter limit=4.00,line
      width=0.800pt] (399.0103,45.7452) .. controls (399.0103,47.6978) and
      (397.4273,49.2807) .. (395.4747,49.2807) .. controls (393.5221,49.2807) and
      (391.9392,47.6978) .. (391.9392,45.7452) .. controls (391.9392,43.7925) and
      (393.5221,42.2096) .. (395.4747,42.2096) .. controls (397.4273,42.2096) and
      (399.0103,43.7925) .. (399.0103,45.7452) -- cycle;
    \path[shift={(-53.03302,-22.72843)},draw=black,fill=cffffff,miter
      limit=4.00,line width=0.800pt] (399.0103,45.7452) .. controls
      (399.0103,47.6978) and (397.4273,49.2807) .. (395.4747,49.2807) .. controls
      (393.5221,49.2807) and (391.9392,47.6978) .. (391.9392,45.7452) .. controls
      (391.9392,43.7925) and (393.5221,42.2096) .. (395.4747,42.2096) .. controls
      (397.4273,42.2096) and (399.0103,43.7925) .. (399.0103,45.7452) -- cycle;
    \path[shift={(-137.88583,-0.50508)},draw=black,fill=cffffff,miter
      limit=4.00,line width=0.800pt] (399.0103,45.7452) .. controls
      (399.0103,47.6978) and (397.4273,49.2807) .. (395.4747,49.2807) .. controls
      (393.5221,49.2807) and (391.9392,47.6978) .. (391.9392,45.7452) .. controls
      (391.9392,43.7925) and (393.5221,42.2096) .. (395.4747,42.2096) .. controls
      (397.4273,42.2096) and (399.0103,43.7925) .. (399.0103,45.7452) -- cycle;
    \path[shift={(-124.75385,48.9924)},draw=black,fill=cffffff,miter limit=4.00,line
      width=0.800pt] (399.0103,45.7452) .. controls (399.0103,47.6978) and
      (397.4273,49.2807) .. (395.4747,49.2807) .. controls (393.5221,49.2807) and
      (391.9392,47.6978) .. (391.9392,45.7452) .. controls (391.9392,43.7925) and
      (393.5221,42.2096) .. (395.4747,42.2096) .. controls (397.4273,42.2096) and
      (399.0103,43.7925) .. (399.0103,45.7452) -- cycle;
    \path[shift={(113.64215,81.31728)},draw=black,fill=cffffff,miter limit=4.00,line
      width=0.800pt] (399.0103,45.7452) .. controls (399.0103,47.6978) and
      (397.4273,49.2807) .. (395.4747,49.2807) .. controls (393.5221,49.2807) and
      (391.9392,47.6978) .. (391.9392,45.7452) .. controls (391.9392,43.7925) and
      (393.5221,42.2096) .. (395.4747,42.2096) .. controls (397.4273,42.2096) and
      (399.0103,43.7925) .. (399.0103,45.7452) -- cycle;
    \path[shift={(102.53047,45.96194)},draw=black,fill=cffffff,miter limit=4.00,line
      width=0.800pt] (399.0103,45.7452) .. controls (399.0103,47.6978) and
      (397.4273,49.2807) .. (395.4747,49.2807) .. controls (393.5221,49.2807) and
      (391.9392,47.6978) .. (391.9392,45.7452) .. controls (391.9392,43.7925) and
      (393.5221,42.2096) .. (395.4747,42.2096) .. controls (397.4273,42.2096) and
      (399.0103,43.7925) .. (399.0103,45.7452) -- cycle;
    \path[shift={(-60.10409,95.45942)},draw=black,fill=cffffff,miter limit=4.00,line
      width=0.800pt] (399.0103,45.7452) .. controls (399.0103,47.6978) and
      (397.4273,49.2807) .. (395.4747,49.2807) .. controls (393.5221,49.2807) and
      (391.9392,47.6978) .. (391.9392,45.7452) .. controls (391.9392,43.7925) and
      (393.5221,42.2096) .. (395.4747,42.2096) .. controls (397.4273,42.2096) and
      (399.0103,43.7925) .. (399.0103,45.7452) -- cycle;
  \end{scope}
  \path[draw=black,line join=miter,line cap=butt,line width=0.480pt]
    (48.2873,210.5448) .. controls (74.9594,206.5036) and (101.6314,206.5036) ..
    (128.3034,210.5448) .. controls (180.5027,222.8713) and (224.6837,211.1432) ..
    (272.5747,210.5448) .. controls (302.3680,202.4576) and (331.0775,204.1241) ..
    (359.2587,210.5448) .. controls (378.5233,219.8884) and (391.8369,213.3626) ..
    (406.5409,210.5448);
  \path[draw=black,line join=miter,line cap=butt,line width=0.480pt]
    (48.2873,236.0045) .. controls (62.1468,231.6454) and (77.3497,229.1671) ..
    (99.2066,236.0045) .. controls (122.0048,245.3293) and (150.3676,244.4520) ..
    (182.8598,236.0045) .. controls (225.7239,236.0591) and (276.2779,251.4932) ..
    (311.3704,236.0045) .. controls (352.9980,247.3235) and (375.0874,236.3131) ..
    (406.5409,236.0045);
  \path[fill=black] (190.05081,234.54469) node[above right] (text4098) {Heisenberg
    Cut};
  \path[fill=black] (61.619301,222.46349) node[above right] (text4102) {?};
  \path[fill=black] (91.593994,227.66312) node[above right] (text4102-2) {?};
  \path[fill=black] (134.0204,231.7037) node[above right] (text4102-7) {?};
  \path[fill=black] (170.38589,229.6631) node[above right] (text4102-21) {?};
  \path[fill=black] (305.74631,224.63266) node[above right] (text4102-8) {?};
  \path[fill=black] (336.05087,229.68343) node[above right] (text4102-5) {?};
  \path[fill=black] (358.27426,226.65292) node[above right] (text4102-82) {?};
  \path[fill=black] (379.48743,229.6834) node[above right] (text4102-0) {?};
  \begin{scope}[cm={{0.51568,0.3069,-0.3069,0.51568,(204.20183,-46.49847)}}]
    \path[draw=black,miter limit=4.00,line width=0.800pt] (291.9341,454.3519) ..
      controls (291.9341,458.2571) and (288.7683,461.4229) .. (284.8630,461.4229) ..
      controls (280.9578,461.4229) and (277.7919,458.2571) .. (277.7919,454.3519) ..
      controls (277.7919,450.4466) and (280.9578,447.2808) .. (284.8630,447.2808) ..
      controls (288.7683,447.2808) and (291.9341,450.4466) .. (291.9341,454.3519) --
      cycle;
    \path[shift={(-1.01015,-4.0)},draw=black,miter limit=4.00,line width=0.800pt]
      (311.1270,458.8976) .. controls (311.1270,463.6396) and (299.8205,467.4839) ..
      (285.8732,467.4839) .. controls (271.9259,467.4839) and (260.6194,463.6396) ..
      (260.6194,458.8976) .. controls (260.6194,454.1555) and (271.9259,450.3113) ..
      (285.8732,450.3113) .. controls (299.8205,450.3113) and (311.1270,454.1555) ..
      (311.1270,458.8976) -- cycle;
    \path[cm={{0.79622,0.60501,-0.60501,0.79622,(335.89171,-82.97669)}},draw=black,miter
      limit=4.00,line width=0.800pt] (311.1270,458.8976) .. controls
      (311.1270,463.6396) and (299.8205,467.4839) .. (285.8732,467.4839) .. controls
      (271.9259,467.4839) and (260.6194,463.6396) .. (260.6194,458.8976) .. controls
      (260.6194,454.1555) and (271.9259,450.3113) .. (285.8732,450.3113) .. controls
      (299.8205,450.3113) and (311.1270,454.1555) .. (311.1270,458.8976) -- cycle;
    \path[cm={{0.83229,-0.55435,0.55435,0.83229,(-206.44342,230.8909)}},draw=black,miter
      limit=4.00,line width=0.800pt] (311.1270,458.8976) .. controls
      (311.1270,463.6396) and (299.8205,467.4839) .. (285.8732,467.4839) .. controls
      (271.9259,467.4839) and (260.6194,463.6396) .. (260.6194,458.8976) .. controls
      (260.6194,454.1555) and (271.9259,450.3113) .. (285.8732,450.3113) .. controls
      (299.8205,450.3113) and (311.1270,454.1555) .. (311.1270,458.8976) -- cycle;
  \end{scope}
  \begin{scope}[cm={{0.53951,-0.26275,0.26275,0.53951,(66.11385,96.63684)}}]
    \path[draw=black,miter limit=4.00,line width=0.800pt] (291.9341,454.3519) ..
      controls (291.9341,458.2571) and (288.7683,461.4229) .. (284.8630,461.4229) ..
      controls (280.9578,461.4229) and (277.7919,458.2571) .. (277.7919,454.3519) ..
      controls (277.7919,450.4466) and (280.9578,447.2808) .. (284.8630,447.2808) ..
      controls (288.7683,447.2808) and (291.9341,450.4466) .. (291.9341,454.3519) --
      cycle;
    \path[shift={(-1.01015,-4.0)},draw=black,miter limit=4.00,line width=0.800pt]
      (311.1270,458.8976) .. controls (311.1270,463.6396) and (299.8205,467.4839) ..
      (285.8732,467.4839) .. controls (271.9259,467.4839) and (260.6194,463.6396) ..
      (260.6194,458.8976) .. controls (260.6194,454.1555) and (271.9259,450.3113) ..
      (285.8732,450.3113) .. controls (299.8205,450.3113) and (311.1270,454.1555) ..
      (311.1270,458.8976) -- cycle;
    \path[cm={{0.79622,0.60501,-0.60501,0.79622,(335.89171,-82.97669)}},draw=black,miter
      limit=4.00,line width=0.800pt] (311.1270,458.8976) .. controls
      (311.1270,463.6396) and (299.8205,467.4839) .. (285.8732,467.4839) .. controls
      (271.9259,467.4839) and (260.6194,463.6396) .. (260.6194,458.8976) .. controls
      (260.6194,454.1555) and (271.9259,450.3113) .. (285.8732,450.3113) .. controls
      (299.8205,450.3113) and (311.1270,454.1555) .. (311.1270,458.8976) -- cycle;
    \path[cm={{0.83229,-0.55435,0.55435,0.83229,(-206.44342,230.8909)}},draw=black,miter
      limit=4.00,line width=0.800pt] (311.1270,458.8976) .. controls
      (311.1270,463.6396) and (299.8205,467.4839) .. (285.8732,467.4839) .. controls
      (271.9259,467.4839) and (260.6194,463.6396) .. (260.6194,458.8976) .. controls
      (260.6194,454.1555) and (271.9259,450.3113) .. (285.8732,450.3113) .. controls
      (299.8205,450.3113) and (311.1270,454.1555) .. (311.1270,458.8976) -- cycle;
  \end{scope}
  \begin{scope}[cm={{0.60009,0.0,0.0,0.60009,(-31.12251,-9.03108)}}]
    \path[draw=black,miter limit=4.00,line width=0.800pt] (291.9341,454.3519) ..
      controls (291.9341,458.2571) and (288.7683,461.4229) .. (284.8630,461.4229) ..
      controls (280.9578,461.4229) and (277.7919,458.2571) .. (277.7919,454.3519) ..
      controls (277.7919,450.4466) and (280.9578,447.2808) .. (284.8630,447.2808) ..
      controls (288.7683,447.2808) and (291.9341,450.4466) .. (291.9341,454.3519) --
      cycle;
    \path[shift={(-1.01015,-4.0)},draw=black,miter limit=4.00,line width=0.800pt]
      (311.1270,458.8976) .. controls (311.1270,463.6396) and (299.8205,467.4839) ..
      (285.8732,467.4839) .. controls (271.9259,467.4839) and (260.6194,463.6396) ..
      (260.6194,458.8976) .. controls (260.6194,454.1555) and (271.9259,450.3113) ..
      (285.8732,450.3113) .. controls (299.8205,450.3113) and (311.1270,454.1555) ..
      (311.1270,458.8976) -- cycle;
    \path[cm={{0.79622,0.60501,-0.60501,0.79622,(335.89171,-82.97669)}},draw=black,miter
      limit=4.00,line width=0.800pt] (311.1270,458.8976) .. controls
      (311.1270,463.6396) and (299.8205,467.4839) .. (285.8732,467.4839) .. controls
      (271.9259,467.4839) and (260.6194,463.6396) .. (260.6194,458.8976) .. controls
      (260.6194,454.1555) and (271.9259,450.3113) .. (285.8732,450.3113) .. controls
      (299.8205,450.3113) and (311.1270,454.1555) .. (311.1270,458.8976) -- cycle;
    \path[cm={{0.83229,-0.55435,0.55435,0.83229,(-206.44342,230.8909)}},draw=black,miter
      limit=4.00,line width=0.800pt] (311.1270,458.8976) .. controls
      (311.1270,463.6396) and (299.8205,467.4839) .. (285.8732,467.4839) .. controls
      (271.9259,467.4839) and (260.6194,463.6396) .. (260.6194,458.8976) .. controls
      (260.6194,454.1555) and (271.9259,450.3113) .. (285.8732,450.3113) .. controls
      (299.8205,450.3113) and (311.1270,454.1555) .. (311.1270,458.8976) -- cycle;
  \end{scope}
  \begin{scope}[cm={{0.60009,0.0,0.0,0.60009,(85.46847,-28.43072)}}]
    \path[draw=black,line join=miter,line cap=butt,line width=0.800pt]
      (339.4113,533.1438) -- (385.8783,568.4991) -- (342.4417,600.8240);
    \path[draw=black,line join=miter,line cap=butt,line width=0.800pt]
      (384.8681,567.4890) .. controls (391.1085,570.3478) and (398.0143,574.8703) ..
      (399.0691,564.7655) .. controls (402.4175,553.8822) and (409.1672,560.0058) ..
      (415.0319,561.7041) .. controls (419.8901,569.6212) and (424.7484,564.1063) ..
      (429.6066,558.9090) .. controls (430.4308,548.3951) and (437.7022,553.9993) ..
      (443.6208,556.2213) .. controls (446.7361,568.4291) and (452.9224,559.1400) ..
      (458.6093,553.3468) -- (488.9138,510.9204);
    \path[draw=black,line join=miter,line cap=butt,line width=0.800pt]
      (458.6093,554.3570) -- (486.8935,592.7428);
  \end{scope}
  \begin{scope}[cm={{0.60009,0.0,0.0,0.60009,(55.56152,-29.64127)}}]
    \path[draw=black,line join=miter,line cap=butt,line width=0.800pt]
      (82.8325,529.1032) -- (116.1675,562.4382) .. controls (119.5725,567.4921) and
      (127.1823,572.5459) .. (115.3252,577.5998) -- (105.5563,588.7861) --
      (114.1472,598.8037) -- (78.7919,620.0169);
    \path[draw=black,line join=miter,line cap=butt,line width=0.800pt]
      (113.1371,597.7935) -- (154.5533,625.0677);
    \path[draw=black,line join=miter,line cap=butt,line width=0.800pt]
      (116.1675,562.4382) -- (167.6853,536.1742);
  \end{scope}
  \path[cm={{0.97255,0.23269,-0.23269,0.97255,(0.0,0.0)}},fill=black]
    (125.96268,237.67122) node[above right] (text4403) {$\Psi$};
  \path[cm={{0.97053,-0.24098,0.24098,0.97053,(0.0,0.0)}},fill=black]
    (191.92426,329.3291) node[above right] (text4403-4) {$\Psi$};
  \path[cm={{0.99133,-0.13138,0.13138,0.99133,(0.0,0.0)}},fill=black]
    (151.6591,340.97943) node[above right] (text4403-7) {$\Psi$};
\end{scope}

\end{tikzpicture}
}
\par\end{center}
\par\end{centering}

\caption{\label{fig:HeisenbergCut}The Heisenberg cut.}
\end{figure}
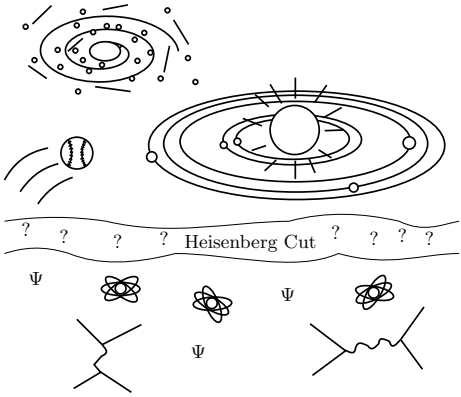

The Copenhagen interpretation is still, at least according to some
surveys \cite{Tegmark:1998iqmmwmw,SchlosshauerKoflerZeilinger:2013sfatqm},
the most popular interpretation of quantum theory today, but it also
suffers from a number of serious drawbacks. Most significantly, the
definition of a measurement according to the Copenhagen interpretation
relies on a questionable demarcation, known as the Heisenberg cut
(\emph{Heisenbergscher Schnitt}) \cite{vonNeumann:1932mgdq,Landsman:2007bcq},
between the large classical systems that carry out measurements and
the small quantum systems that they measure; this ill-defined Heisenberg
cut has never been identified in any experiment to date and must be
put into the interpretation by hand. (See Figure~\ref{fig:HeisenbergCut}.)
An associated issue is the interpretation's assumption of wave-function collapse,
by which we refer to the supposed instantaneous, discontinuous change
in a quantum system immediately following a measurement by a classical
system, in stark contrast to the smooth time evolution that governs
dynamically closed systems.

The Copenhagen interpretation is also unclear as to the ultimate meaning
of the state vector of a system: Does a system's state vector merely
represent the experimenter's knowledge, is it some sort of objective
probability distribution,%
\footnote{Recent work \cite{PuseyBarrettRudolph:2012rqs,BarrettCavalcantiLalMaroney:2013rqs,ColbeckRenner:2013swfudups}
casts considerable doubt on assertions that state vectors are nothing
more than probability distributions over more fundamental ingredients
of reality.%
} or is it an irreducible ingredient of reality like the state of a
classical system? For that matter, what constitutes an observer,
and can we meaningfully talk about the state of an observer within
the formalism of quantum theory? Given that no realistic system is
ever perfectly free of quantum entanglements with other systems, and
thus no realistic system can ever truly be assigned a specific state
vector in the first place, what becomes of the popular depiction of
quantum theory in which every particle is supposedly described by
a specific wave function propagating in three-dimensional space? The
Copenhagen interpretation leads to additional trouble when trying
to make sense of thought experiments like Schrödinger's cat, Wigner's friend,
and the quantum Zeno paradox.%
\footnote{We discuss all of these thought experiments in our companion paper
\cite{BarandesKagan:2014mmiqt}.%
}

A more satisfactory interpretation would eliminate the need for an
\emph{ad hoc} Heisenberg cut, thereby demoting measurements to an
ordinary kind of interaction and allowing quantum theory to be a complete
theory that seamlessly encompasses \emph{all} systems in Nature, including
observers as physical systems with quantum states of their own. Moreover,
such an interpretation should fundamentally (even if not always \emph{superficially})
be consistent with all experimental data and other reliably known
features of Nature, including relativity, and should be general enough
to accommodate the large variety of both presently known and hypothetical
physical systems. Such an interpretation should also address the key
no-go theorems developed over the years by researchers working on
the foundations of quantum theory, should not depend on concepts or
quantities whose definitions require a physically unrealistic measure-zero
level of sharpness, and should be insensitive to potentially unknowable
features of reality, such as whether we can sensibly define ``the
universe as a whole'' as being a closed or open system.

\subsection{Motivation from Classical Physics\label{sub:Motivation-from-Classical-Physics}}

Our interpretation is fully quantum in nature. However, for purposes
of motivation, consider the basic theoretical structure of classical
physics: A classical system has a specific state---which we call its
``ontic state''---that evolves in time through the system's configuration
space according to some dynamical rule that may or may not be stochastic,
and this dynamical rule exists whether or not the system's state lies
beneath a nontrivial evolving probability distribution---called an
``epistemic state''---on the system's configuration space; moreover,
the dynamical rule for the system's underlying ontic state is consistent
with the overall evolution of the system's epistemic state, in the
sense that if we consider a probabilistic ensemble over the system's
initial ontic state and apply the dynamical rule to each ontic state
in the ensemble, then we correctly obtain the overall evolution of
the system's epistemic state as a whole.

In particular, note the \emph{insufficiency} of specifying a dynamical
rule \emph{solely} for the evolution of the system's overall probability
distribution---that is, for its epistemic state alone---but \emph{not}
for the system's underlying ontic state itself, because then the system's
underlying ontic state would be free to fluctuate wildly and discontinuously
between macroscopically distinct configurations. For example, even
if a classical system's epistemic state describes constant probabilities
$p_{1}$ and $p_{2}$ for the system to be in macroscopically distinct
ontic states $q_{1}$ or $q_{2}$, there would be nothing preventing
the system's ontic state from hopping discontinuously between $q_{1}$
and $q_{2}$ with respective frequency ratios $p_{1}$ and $p_{2}$
over arbitrarily short time intervals. Essentially, by imposing a
dynamical rule on the system's underlying ontic state, we can provide
a ``smoothness condition'' for the system's physical configuration
over time and thus eliminate these kinds of instabilities.

In quantum theory, a system that is \emph{exactly} closed and that
is \emph{exactly} in a pure state (both conditions that are unphysical
idealizations) evolves along a well-defined trajectory through the
system's Hilbert space according to a well-known dynamical rule, namely,
the Schrödinger equation. However, in traditional formulations of
quantum theory, an open quantum system that must be described by a
density matrix due to entanglements with other systems---a so-called
improper mixture---does not have a specific underlying ontic state
vector, let alone a Hilbert-space trajectory or a dynamical rule governing
the time evolution of such an underlying ontic state vector and consistent
with the overall evolution of the system's density matrix. It is a
chief goal of our interpretation of quantum theory to provide these
missing ingredients.

\section{The Minimal Modal Interpretation}

\subsection{Conceptual Summary\label{sub:Conceptual-Summary}}

For a quantum system in an improper mixture, our new interpretation
identifies the \emph{eigenstates} of the system's density matrix with
the \emph{possible states} of the system in reality and identifies
the \emph{eigenvalues} of that density matrix with the \emph{probabilities}
that one of those possible states is \emph{actually} occupied, and
introduces just enough minimal structure beyond this simple picture
to provide a dynamical rule for underlying state vectors as they evolve
along Hilbert-space trajectories and to evade criticisms made in the
past regarding similar interpretations. This minimal additional structure
consists of a single additional class of conditional probabilities
amounting essentially to a series of smoothness conditions that kinematically
relate the states of parent systems to the states of their subsystems,
as well as dynamically relate the states of a single system to each
other over time.

\subsection{Technical Summary\label{sub:Technical-Summary}}

\begin{figure}
\begin{centering}
\begin{center}
\scalebox{1.5}{
\definecolor{ce1e1e1}{RGB}{225,225,225}

\begin{tikzpicture}[y=0.80pt, x=0.8pt,yscale=-1, inner sep=0pt, outer sep=0pt]
\begin{scope}[shift={(-41.15005,-13.21949)}]
  \path[draw=black,fill=ce1e1e1,line join=miter,miter limit=4.00,line width=0.761pt,rounded
    corners=0.2000cm] (165,51) rectangle (182,70);
  \path[fill=black] (92.06411,87.079956) node[above right] (text5199-52)
    {$\hat{\rho}$};
  \path[->,>=latex,draw=black,line join=miter,line cap=butt,line width=0.924pt]
    (108.6989,83.3966) -- (143.0093,83.3966);
  \path[->,>=latex,draw=black,line join=miter,line cap=butt,line width=0.925pt]
    (109.0520,75) -- (142.3517,63.2528);
  \path[->,>=latex,draw=black,line join=miter,line cap=butt,line width=1.015pt]
    (107.9561,67.1956) -- (141.6060,42.5951);
  \path[fill=black] (120.57199,103.81787) node[above right] (text5199-52-2)
    {$\vdots$};
  \path[fill=black] (150.37277,46.663536) node[above right] (text5199-52-1) {$p_1,
    \Psi_1$};
  \path[fill=black] (150.37251,66.615074) node[above right] (text5199-52-1-9)
    {$p_2, \Psi_2$};
  \path[fill=black] (150.37251,87.630531) node[above right] (text5199-52-1-9-5)
    {$p_3, \Psi_3$};
  \path[->,>=latex,draw=black,line join=miter,line cap=butt,line width=0.924pt]
    (140,120) -- (152,95);
  \path[->,>=latex,draw=black,line join=miter,line cap=butt,line width=0.924pt]
    (172,120) -- (172,95);
  \path[->,>=latex,draw=black,line join=miter,line cap=butt,line width=0.924pt]
    (223,60) -- (192,60);
\path[fill=black] (100,145) node[above right] (text5199-52-2-0) {$\begin{subarray}{l}\text{epistemic} \\ \text{probabilities} \\ \text{(eigenvalues)}\end{subarray}$};
\path[fill=black] (170,143) node[above right] (text5199-52-2-0) {$\begin{subarray}{l}\text{possible} \\ \text{ontic states} \\ \text{(eigenstates)}\end{subarray}$};
\path[fill=black] (230,65) node[above right] (text5199-52-2-0) {$\begin{subarray}{l}\text{actual} \\ \text{ontic state} \end{subarray}$};
\end{scope}

\end{tikzpicture}
}
\par\end{center}
\par\end{centering}

\caption{\label{fig:DensityMatrixFulcrum}A schematic depiction of our postulated
relationship between a system's density matrix $\op{\rho}$ and its
associated epistemic state $\set{\left(p_{1},\Psi_{1}\right),\left(p_{2},\Psi_{2}\right),\left(p_{3},\Psi_{3}\right),\dotsc}$,
the latter consisting of epistemic probabilities $p_{1},p_{2},p_{3},\dotsc$
(the eigenvalues of the density matrix) and possible ontic states
$\Psi_{1},\Psi_{2},\Psi_{3},\dotsc$ (represented by the eigenstates
of the density matrix), where one of those \emph{possible} ontic states
(in this example, $\Psi_{2}$) is the system's \emph{actual} ontic
state.}
\end{figure}
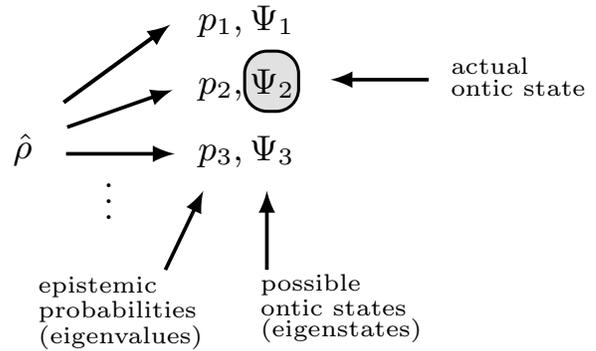

Our new interpretation, which we call the \emph{minimal modal interpretation}
of quantum theory and which we motivate and detail more extensively
in our companion paper \cite{BarandesKagan:2014mmiqt}, consists of
several parsimonious ingredients:
\begin{enumerate}
\item We define ontic states $\Psi_{i}$, meaning the states of the given
system as it could actually exist in reality, in terms of arbitrary
(unit-norm) state vectors $\ket{\Psi_{i}}$ in the system's Hilbert
space $\mathcal{H}$, 
\begin{equation}
\Psi_{i}\exchange\ket{\Psi_{i}}\in\mathcal{H}\ \left(\mathrm{up\ to\ overall\ phase}\right),\label{eq:OntStateVecCorresp}
\end{equation}
 and we define epistemic states $\setindexed{\left(p_{i},\Psi_{i}\right)}i$
as probability distributions over sets of possible ontic states, 
\[
\setindexed{\left(p_{i},\Psi_{i}\right)}i,\qquad p_{i}\in\left[0,1\right],
\]
 where these definitions parallel the corresponding notions from classical
physics. We translate logical mutual exclusivity of ontic states $\Psi_{i}$
as mutual orthogonality of state vectors $\ket{\Psi_{i}}$, and we
make a distinction between subjective epistemic states (proper mixtures)
and objective epistemic states (improper mixtures): The former arise
from classical ignorance and are uncontroversial, whereas the latter
arise from quantum entanglements to other systems and do not have
a widely accepted \emph{a priori} meaning outside of our interpretation
of quantum theory. Indeed, the problem of interpreting objective epistemic
states may well be unavoidable: Essentially all realistic systems
are entangled to other systems to a nonzero degree and thus cannot
be described exactly by pure states or by purely subjective epistemic
states.%
\footnote{As we explain in Section~\ref{sub:Comparison-with-Other-Interpretations-of-Quantum-Theory},
there are reasons to be skeptical of the common assumption that one
can always assign an exact pure state or purely subjective epistemic
state to ``the universe as a whole.''%
}
\item We posit a correspondence between objective epistemic states $\setindexed{\left(p_{i},\Psi_{i}\right)}i$
and density matrices $\op{\rho}$: 
\begin{equation}
\setindexed{\left(p_{i},\Psi_{i}\right)}i\exchange\op{\rho}=\sum_{i}p_{i}\ket{\Psi_{i}}\bra{\Psi_{i}}.\label{eq:EpStateDensMatrixCorresp}
\end{equation}
 (See Figure~\ref{fig:DensityMatrixFulcrum}.) The relationship between
subjective epistemic states and density matrices is not as strict,
as we explain in our companion paper \cite{BarandesKagan:2014mmiqt}.

\item We invoke the partial-trace operation $\op{\rho}_{Q}\equiv\Tr_{E}\left[\op{\rho}_{Q+E}\right]$,
motivated and defined in our companion paper \cite{BarandesKagan:2014mmiqt}
without appeals to the Born rule or Born-rule-based averages, to relate
the density matrix (and thus the epistemic state) of any subsystem
$Q$ to that of any parent system $W=Q+E$.
\item We introduce a general class of quantum conditional probabilities,
\begin{equation}
\begin{aligned} & \negthickspace\negthickspace\negthickspace\negthickspace p_{Q_{1},\dotsc,Q_{n}\given W}\left(i_{1},\dotsc,i_{n};t^{\prime}\given w;t\right)\\
 & \negthickspace\negthickspace\negthickspace\negthickspace\equiv\Tr_{W}\left[\left(\op P_{Q_{1}}\left(i_{1};t^{\prime}\right)\tensorprod\dotsm\tensorprod\op P_{Q_{n}}\left(i_{n};t^{\prime}\right)\right)\mathcal{E}_{W}^{t^{\prime}\from t}\left[\op P_{W}\left(w;t\right)\right]\right]\\
 & \sim\Tr\left[\op P_{i_{1}}\left(t^{\prime}\right)\dotsm\op P_{i_{n}}\left(t^{\prime}\right)\mathcal{E}\left[\op P_{w}\left(t\right)\right]\right],
\end{aligned}
\label{eq:DefQuantCondProbGenFromTrParent}
\end{equation}
 relating the possible ontic states of any partitioning collection
of mutually disjoint subsystems $Q_{1},\dotsc,Q_{n}$ to the possible
ontic states of a corresponding parent system $W=Q_{1}+\dotsb+Q_{n}$
whose own dynamics is governed by a linear completely-positive-trace-preserving
(``CPT'') dynamical mapping $\mathcal{E}_{W}^{t^{\prime}\from t}\left[\cdot\right]$
over the given time interval $t^{\prime}-t$.  Here $\op P_{W}\left(w;t\right)$
denotes the projection operator onto the eigenstate $\ket{\Psi_{W}\left(w;t\right)}$
of the density matrix $\op{\rho}_{W}\left(t\right)$ of the parent
system $W$ at the initial time $t$, and, similarly, $\op P_{Q_{\alpha}}\left(i;t^{\prime}\right)$
denotes the projection operator onto the eigenstate $\ket{\Psi_{Q_{\alpha}}\left(i;t^{\prime}\right)}$
of the density matrix $\op{\rho}_{Q_{\alpha}}\left(t^{\prime}\right)$
of the subsystem $Q_{\alpha}$ at the final time $t^{\prime}$ for
$\alpha=1,\dotsc,n$. In a rough sense, the dynamical mapping $\mathcal{E}_{W}^{t^{\prime}\from t}\left[\cdot\right]$
acts as a parallel-transport superoperator that moves the parent-system
projection operator $\op P_{W}\left(w;t\right)$ from $t$ to $t^{\prime}$
before we compare it with the subsystem projection operators $\op P_{Q_{\alpha}}\left(i;t^{\prime}\right)$.\\
\\
In generalizing unitary dynamics to linear CPT dynamics in this manner,
as is necessary in order to account for the crucial and non-reductive
quantum relationships between parent systems and their subsystems,
note that we are \emph{not} proposing any fundamental modification
to the dynamics of quantum theory, such as in GRW-type spontaneous-localization
models \cite{GhirardiRiminiWeber:1986udmms}, but are simply accommodating
the fact that generic mesoscopic and macroscopic quantum systems are
typically open to their environments to some nonzero degree. Indeed,
linear CPT dynamical mappings are widely used in quantum chemistry
as well as in quantum information science, in which they are known
as quantum operations; when specifically regarded as carriers of
quantum information, they are usually called\emph{ }quantum channels.%
\footnote{See \cite{SudarshanMatthewsRau:1961sdqms,JordanSudarshan:1961dmdoqm,Choi:1975cplmcm}
for early work in this direction, and see \cite{SchumacherWestmoreland:2010qpsi}
for a modern pedagogical review. Starting from a simple measure of
distinguishability between density matrices that is non-increasing
under linear CPT dynamics \cite{Ruskai:1994bss}, one can argue \cite{BreuerLainePiilo:2009mdnmbqpos,LainePiiloBreuer:2010mnmqp}
that linear CPT dynamics implies the nonexistence of the backward
flow of information into the system from its environment. \cite{Buscemi:2013cppic}
strengthens this reasoning by proving that exact linear CPT dynamics
exists for a given quantum system if and only if the system's initial
correlations with its environment satisfy a quantum data-processing
inequality that prevents backward information flow.%
} A well-known, concrete example is the Lindblad equation \cite{Lindblad:1976gqds}.\\
\\
As special cases, these quantum conditional probabilities provide
a \emph{kinematical} smoothing relationship 
\begin{equation}
\begin{aligned} & p_{Q_{1},\dotsc,Q_{n}\given W}\left(i_{1},\dotsc,i_{n}\given w\right)\\
 & \equiv\Tr_{W}\left[\left(\op P_{Q_{1}}\left(i_{1}\right)\tensorprod\dotsm\tensorprod\op P_{Q_{n}}\left(i_{n}\right)\right)\op P_{W}\left(w\right)\right]\\
 & =\bra{\Psi_{W,w}}\left(\ket{\Psi_{Q_{1},i_{1}}}\bra{\Psi_{Q_{1},i_{1}}}\tensorprod\dotsm\right.\\
 & \qquad\left.\tensorprod\ket{\Psi_{Q_{n},i_{n}}}\bra{\Psi_{Q_{n},i_{n}}}\right)\ket{\Psi_{W,w}}
\end{aligned}
\label{eq:QuantCondProbKinParentSub}
\end{equation}
 between the possible ontic states of any partitioning collection
of mutually disjoint subsystems $Q_{1},\dotsc,Q_{n}$ and the possible
ontic states of the corresponding parent system $W=Q_{1}+\dotsb+Q_{n}$
at any \emph{single} moment in time, and, taking $Q\equiv Q_{1}=W$,
also provide a \emph{dynamical} smoothing relationship 
\begin{equation}
\begin{aligned}p_{Q}\left(j;t^{\prime}\given i;t\right) & \equiv\Tr_{Q}\left[\op P_{Q}\left(j;t^{\prime}\right)\mathcal{E}_{Q}^{t^{\prime}\from t}\left[\op P_{Q}\left(i;t\right)\right]\right]\\
 & \sim\Tr\left[\op P_{j}\left(t^{\prime}\right)\mathcal{E}\left[\op P_{i}\left(t\right)\right]\right]
\end{aligned}
\label{eq:ExactDynQuantCondProbs}
\end{equation}
 between the possible ontic states of a system $Q$ over time and
also between the objective epistemic states of a system $Q$ over
time. 
\end{enumerate}
Essentially, 1 establishes a linkage between ontic states and epistemic
states, 2 establishes a linkage between (objective) epistemic states
and density matrices, 3 establishes a linkage between parent-system
density matrices and subsystem density matrices, and 4 establishes
a linkage between parent-system ontic states and subsystem ontic states,
either at the same time or at different times.

After verifying in our companion paper \cite{BarandesKagan:2014mmiqt}
that our quantum conditional probabilities satisfy a number of consistency
requirements, we show that they allow us to avoid ontological instabilities
that have presented problems for other modal interpretations, analyze
the measurement process, study various familiar ``paradoxes'' and
thought experiments, and examine the status of Lorentz invariance
and locality in our interpretation of quantum theory. In particular,
our interpretation accommodates the nonlocality implied by the EPR-Bohm
and GHZ-Mermin thought experiments without leading to superluminal
signaling, and evades claims by Myrvold \cite{Myrvold:2002mir} purporting
to show that interpretations like our own lead to unacceptable contradictions
with Lorentz invariance. As a consequence of its compatibility with
Lorentz invariance, we claim that our interpretation is capable of
encompassing all the familiar quantum models of physical systems widely
in use today, from nonrelativistic point particles to quantum field
theories and even string theory.

\subsection{Modal Interpretations\label{sub:Modal-Interpretations}}

Our interpretation of quantum theory belongs to the general class
of modal interpretations originally introduced by Krips in 1969 \cite{Krips:1969tpqm,Krips:1975siqt,Krips:1987mqt}
and then independently developed by van Fraassen (whose early formulations
involved the fusion of modal logic with quantum logic), Dieks, Vermaas,
and others \cite{vanFraassen:1972faps,Cartwright:1974vfmmqm,VermaasDieks:1995miqmgdo,BacciagaluppiDickson:1999dmi,Vermaas:1999puqm}.

The modal interpretations are now understood to encompass a very large
set of interpretations of quantum theory, including most interpretations
that fall between the ``many worlds'' of the Everett-DeWitt approach
and the ``no worlds'' of the instrumentalist approaches.  Generally
speaking, in a modal interpretation, one singles out some preferred
basis for each system's Hilbert space and then regards the elements
of that basis as the system's \emph{possible} ontic states---one of
which is the system's \emph{actual} ontic state---much in keeping
with how we think conceptually about classical probability distributions.
For example, as emphasized in \cite{Vermaas:1999puqm}, the de Broglie-Bohm
pilot-wave interpretation \cite{deBroglie:1930iswm,Bohm:1952siqtthvi,Bohm:1952siqtthvii}
can be regarded as a special kind of modal interpretation in which
the preferred basis is permanently fixed \emph{for all} systems at
a universal choice. Other modal interpretations, such as our own,
instead allow the preferred basis for a given system to change---in
our case by choosing the preferred basis to be the evolving eigenbasis
of that system's density matrix.  However, we claim that no existing
modal interpretation captures the one that we summarize in this letter
and describe more fully in our companion paper \cite{BarandesKagan:2014mmiqt}.

\subsection{Hidden Variables and the Irreducibility of Ontic States\label{sub:Hidden-Variables-and-the-Irreducibility-of-Ontic-States}}

To the extent that our interpretation of quantum theory involves hidden variables,
the actual ontic states underlying the epistemic states of systems
play that role. However, one could also argue that calling them hidden
variables is just an issue of semantics because they are on the same
metaphysical footing as both the \emph{traditional} notion of quantum
states as well as the actual ontic states of \emph{classical} systems.

In any event, it is important to note that our interpretation includes
\emph{no other} hidden variables: Just as in the classical case, we
regard ontic states as being \emph{irreducible} objects, and, in keeping
with this interpretation, we  \emph{do not} regard a system's ontic
state itself as being an epistemic probability distribution---much
less a ``pilot wave''---over a set of more basic hidden variables.
In a rough sense, our interpretation \emph{unifies} the de Broglie-Bohm
interpretation's pilot wave and hidden variables into a single ontological
entity that we call an ontic state.

In particular, we do not attach an epistemic probability interpretation
to the \emph{components} of a vector representing a system's ontic
state, nor do we assume \emph{a priori} the Born rule, which we 
ultimately \emph{derive} in our companion paper \cite{BarandesKagan:2014mmiqt}
as a means of computing empirical outcome probabilities. Otherwise,
we would need to introduce an unnecessary \emph{additional }level
of probabilities into our interpretation and thereby reduce its axiomatic
parsimony and explanatory power.

Via the phenomenon of environmental decoherence, our interpretation
ensures that the evolving ontic state of a sufficiently macroscopic
system---with significant energy and in contact with a larger environment---is
highly likely to be represented by a temporal sequence of state vectors
whose labels evolve in time according to recognizable semi-classical
equations of motion. For microscopic, isolated systems, by contrast,
we simply accept that the ontic state vector may not always have an
intuitively familiar classical description.

\subsection{Comparison with Other Interpretations of Quantum Theory\label{sub:Comparison-with-Other-Interpretations-of-Quantum-Theory}}

Our interpretation, which builds on the work of many others, is general,
model-independent, and encompasses relativistic systems, but is also
conservative and unextravagant: It includes only metaphysical objects
that are either already a standard part of quantum theory or that
have counterparts in classical physics. We do not posit the existence
of exotic ``many worlds'' \cite{Everett:1957rsfqm,EverettDeWittGraham:1973mwiqm,Deutsch:1985qtupt,Wallace:2002wei},
physical ``pilot waves'' \cite{deBroglie:1930iswm,Bohm:1952siqtthvi,Bohm:1952siqtthvii},
or any fundamental GRW-type dynamical-collapse or spontaneous-localization
modifications to quantum theory \cite{GhirardiRiminiWeber:1986udmms,Pearle:1989csdsvrwsl,BassiGhirardi:2003drm,Weinberg:2011csv}.
Indeed, our interpretation leaves the widely accepted mathematical
structure of quantum theory essentially intact.%
\footnote{Moreover, our interpretation does not introduce any new violations
of time-reversal symmetry, and gives no fundamental role to  relative
states \cite{Everett:1957rsfqm}; a cosmic multiverse or self-locating
uncertainty \cite{AguirreTegmark:2011biuciqm}; coarse-grained histories
or decoherence functionals \cite{Hartle:1991scgnqm,GellMannHartle:2013acgesdqr};
decision theory \cite{Deutsch:1999qtpd,Wallace:2002qpdtr}; Dutch-book
coherence, SIC-POVMs, or \emph{urgleichungs} \cite{FuchsSchack:2013qbcnnv};
circular frequentist arguments involving unphysical ``limits'' of
infinitely many copies of measurement devices \cite{FarhiGoldstoneGutmann:1989hpaqm,BuniyHsuZee:2006dopqm,AguirreTegmark:2011biuciqm};
infinite imaginary ensembles \cite{Ballentine:1970siqm}; quantum
reference systems or perspectivalism \cite{Bene:1997qrsnfqm,BeneDieks:2002pvmiqm};
relational or non-global quantum states \cite{Rovelli:1996rqm,BerkovitzHemmo:2005miqmrr,Hollowood:2013eciqm,Hollowood:2013nmqm};
many-minds states \cite{AlbertLoewer:1988imwi}; mirror states \cite{Hollowood:2013eciqm,Hollowood:2013nmqm};
faux-Boolean algebras \cite{Dickson:1995fbacm,Vermaas:1999puqm};
``atomic'' subsystems \cite{BacciagaluppiDickson:1997gqtm}; algebraic
quantum field theory \cite{EarmanRuetsche:2005rimi}; secret classical
superdeterminism or fundamental information loss \cite{tHooft:2005dbqm};
cellular automata \cite{tHooft:2014caiqm}; classical matrix degrees
of freedom or trace dynamics \cite{Adler:2002sdguimmpqm}; or discrete
Hilbert spaces or appeals to unknown Planck-scale physics \cite{BuniyHsuZee:2006dopqm}.%
} At the same time, we argue in our companion paper \cite{BarandesKagan:2014mmiqt}
that our interpretation is ultimately compatible with Lorentz invariance
and is nonlocal only in the mild sense familiar from the framework
of classical gauge theories.

Furthermore, we make no assumptions about as-yet-unknown aspects of
reality, such as the fundamental discreteness or continuity of time
or the dimensionality of the ultimate Hilbert space of Nature. Nor
does our interpretation rely in any crucial way upon the existence
of a well-defined maximal parent system that encompasses all other
systems and is dynamically closed in the sense of having a so-called
cosmic pure state or universal wave function\emph{ }that \emph{precisely}
obeys the Schrödinger equation; by contrast, this sort of cosmic assumption
is a necessarily \emph{exact} ingredient in the traditional formulations
of the de Broglie-Bohm pilot-wave interpretation and the Everett-DeWitt many-worlds interpretation.

Indeed, by considering merely the \emph{possibility} that our observable
universe is but a small region of an eternally inflating cosmos of
indeterminate spatial size and age \cite{Linde:1986eci,GarrigaVilenkin:1998ru,Guth:2007eii},
it becomes clear that the idea of a biggest closed system (``the
universe as a whole'') may not generally be a sensible or empirically
verifiable concept to begin with, let alone an axiom on which a robust
interpretation of quantum theory can safely rely. Our interpretation
certainly allows for the existence of a biggest closed system, but
is also fully able to accommodate the alternative circumstance that
if we were to imagine gradually enlarging our scope to parent systems
of increasing physical size, then we might well find that the hierarchical
sequence never terminates at any maximal, dynamically closed system,
but may instead lead to an unending ``Russian-doll'' succession
of ever-more-open parent systems.%
\footnote{One might try to argue that one can always \emph{formally} define
a closed maximal parent system just to be ``the system containing
all systems.'' Whatever logicians might say about such a construction,
we run into the more prosaic issue that if we cannot construct this
closed maximal parent system via a well-defined succession of parent
systems of incrementally increasing size, then it becomes unclear
mathematically how we can generally define any human-scale system
as a subsystem of the maximal parent system and thereby define the
partial-trace operation. Furthermore, if our observable cosmic region
is indeed an open system, then its own time evolution may not be exactly
linear, in which case it is far from obvious that we can safely and
rigorously embed that open-system time evolution into the hypothetical
unitary dynamics of any conceivable closed parent system.%
}

\section{Measurements and Lorentz Invariance\label{sec:Measurements-and-Lorentz-Invariance}}

\subsection{Von Neumann Measurements and the Born Rule\label{sub:Von-Neumann-Measurements-and-the-Born-Rule}}

For the purposes of establishing how our minimal modal interpretation
makes sense of measurements, how decoherence turns the environment
into a ``many-dimensional chisel'' that rapidly sculpts the ontic
states of systems into its precise shape,%
\footnote{The way that decoherence sculpts ontic states into shape is reminiscent
of the way that the external pressure of air molecules above a basin
of water maintain the water in its liquid phase.%
} and how the Born rule naturally emerges to an excellent approximation,
we consider in our companion paper \cite{BarandesKagan:2014mmiqt}
the idealized example of a so-called Von Neumann measurement. Along
the way, we also address the status of both the measurement problem
generally and the notion of wave-function collapse specifically in
the context of our interpretation of quantum theory. Ultimately, we
find that our interpretation solves the measurement problem by replacing
instantaneous axiomatic wave-function collapse with an interpolating
ontic-level dynamics, and thereby eliminates any need for a Heisenberg
cut.

\subsection{The Myrvold No-Go Theorem\label{sub:The-Myrvold-No-Go-Theorem}}

Building on a paper by Dickson and Clifton \cite{DicksonClifton:1998limi}
and employing arguments similar to Hardy \cite{Hardy:1992qmlrtlirt},
Myrvold \cite{Myrvold:2002mir} argued that modal interpretations
are fundamentally inconsistent with Lorentz invariance at a deeper
level than mere unobservable nonlocality, leading to much additional
work \cite{EarmanRuetsche:2005rimi,BerkovitzHemmo:2005miqmrr,Myrvold:2009cc}
in subsequent years to determine the implications of his result. Specifically,
Myrvold argued that one could not safely assume that a quantum system,
regardless of its size or complexity, always possesses a specific
actual ontic state beneath its epistemic state, because any such actual
ontic state could seemingly change under Lorentz transformations more
radically than, say, a four-vector does---for example, binary digits
$1$ and $0$ could switch, words written on a paper could change,
and true could become false. The only way to avoid this conclusion
would then be to break Lorentz symmetry in a fundamental way by asserting
the existence of a ``preferred'' Lorentz frame in which all ontic-state
assignments must be made.

Myrvold's claims, and those of Dickson and Clifton as well as Hardy,
rest on several invalid assumptions. Dickson and Clifton \cite{DicksonClifton:1998limi}
assume that the hidden ontic states admit certain joint epistemic
probabilities that are conditioned on \emph{multiple} disjoint systems
at an initial time, and such probabilities are not a part of our interpretation
of quantum theory, as we explain in our companion paper \cite{BarandesKagan:2014mmiqt}.%
\footnote{The same implicit assumption occurs in Section 9.2 of \cite{Vermaas:1999puqm}.%
} Similarly, one of Myrvold's arguments hinges on the assumed existence
of joint epistemic probabilities for two disjoint systems at \emph{two
separate times}, and again such probabilities are not present in our
interpretation. As Dickson and Clifton point out in Appendix B of
their paper, Hardy's argument also relies on several faulty assumptions
about ontic property assignments in modal interpretations.

A final argument of Myrvold---and repeated by Berkovitz and Hemmo
\cite{BerkovitzHemmo:2005miqmrr}---is based on inadmissible assumptions
about the proper way to implement Lorentz transformations in quantum
theory and the relationship between density matrices and ontic states
in our minimal modal interpretation. Myrvold's mistake is subtle,
so we go through his argument in detail in our companion paper \cite{BarandesKagan:2014mmiqt}.

\subsection{Quantum Theory and Classical Gauge Theories\label{sub:Quantum-Theory-and-Classical-Gauge-Theories}}

Suppose that we were to imagine reifying all of the \emph{possible}
ontic states defined by each system's density matrix as simultaneous
\emph{actual} ontic states in the sense of the many-worlds interpretation.%
\footnote{Observe that the eigenbasis of each system's density matrix therefore
defines a preferred basis \emph{for that system alone}. We \emph{do
not} assume the sort of universe-spanning preferred basis shared by
all systems that is featured in the traditional many-worlds interpretation;
such a universe-spanning preferred basis would lead to new forms of
nonlocality, as we describe in our companion paper \cite{BarandesKagan:2014mmiqt}.%
} Then because every density matrix \emph{as a whole} evolves locally,
no nonlocal dynamics between the actual ontic states is necessary
and our interpretation of quantum theory becomes manifestly dynamically
local: For example, each spin detector in the EPR-Bohm or GHZ-Mermin
thought experiments possesses all its possible results in actuality,
and the larger measurement apparatus locally ``splits'' into all
the various possibilities when it visits each spin detector and looks
at the detector's final reading.

In our companion paper \cite{BarandesKagan:2014mmiqt}, we argue that
we can make sense of this step of adding unphysical actual ontic states
into our interpretation of quantum theory by appealing to an analogy
with classical gauge theories, and specifically the example of the
Maxwell theory of electromagnetism. Just as different choices of
gauge for a given classical gauge theory make different calculations
or properties of the theory more or less manifest---each choice of
gauge inevitably involves trade-offs---we see that switching from
the ``unitary gauge'' corresponding to our interpretation of quantum
theory to the ``Lorenz gauge'' in which it looks more like a density-matrix-centered
version of the many-worlds interpretation makes the locality and Lorentz
covariance of the interpretation more manifest at the cost of obscuring
the interpretation's underlying ontology and the meaning of probability.

Seen from this perspective, we can also better understand why it is
so challenging \cite{Hsu:2012oopqm} to make sense of a many-worlds-type
interpretation as an ontologically and epistemologically reasonable
interpretation of quantum theory: Attempting to do so leads to as
much metaphysical difficulty as trying to make sense of the Lorenz
gauge of Maxwell electromagnetism as an ``ontologically correct interpretation''
of the Maxwell theory.%
\footnote{Indeed, in large part for this reason, some textbooks \cite{Weinberg:1996qtfi}
develop quantum electrodynamics fundamentally from the perspective
of Weyl-Coulomb gauge $A^{0}=\divergence\svec A=0$.%
} Hence, taking a lesson from classical gauge theories, we propose
instead regarding many-worlds-type interpretations as merely a convenient
mathematical tool---a particular ``gauge choice''---for establishing
definitively that a given ``unitary-gauge'' interpretation of quantum
theory like our own is ultimately consistent with locality and Lorentz
invariance.

\section{Conclusion\label{sec:Conclusion}}

In this letter and in a more extensive companion paper \cite{BarandesKagan:2014mmiqt},
we have introduced a new, conservative interpretation of quantum theory
that threads a number of key requirements that we feel are insufficiently
addressed by other interpretations. In particular, our interpretation
identifies a definite ontology for every quantum system, as well as
dynamics for that ontology based on the overall time evolution of
the system's density matrix, and sews together ontologies for parent
systems and their subsystems in a natural way.

\subsection*{Falsifiability and the Role of Decoherence}

As some other interpretations do, our own interpretation puts decoherence
in the central role of transforming the Born rule from an axiomatic
postulate into a derived consequence and thereby solving the measurement
problem of quantum theory. We regard it as a positive feature of our
interpretation of quantum theory that falsification of the capacity
for decoherence to manage this responsibility would mean falsification
of our interpretation. We therefore also take great interest in the
ongoing arms race between proponents and critics of decoherence, in
which critics offer up examples of decoherence coming up short \cite{AlbertLoewer:1990wdatassp,AlbertLoewer:1991mpss,AlbertLoewer:1993nim,Albert:1994qme,Page:2011qusbgbd}
and thereby push proponents to argue that increasingly realistic measurement
set-ups involving non-negligible environmental interactions resolve
the claimed inconsistencies \cite{BacciagaluppiHemmo:1996midm,Hollowood:2013cbr,Hollowood:2013eciqm,Vermaas:1999puqm}.

\subsubsection*{ACKNOWLEDGMENTS}
\begin{acknowledgments}
J.\,A.\,B. has benefited tremendously from personal communications
with Matthew Leifer, Timothy Hollowood, and Francesco Buscemi. D.\,K.
thanks Gaurav Khanna, Pontus Ahlqvist, Adam Brown, Darya Krym, and
Paul Cadden-Zimansky for many useful discussions on related topics,
and is supported in part by FQXi minigrant \#FQXi-MGB-1322. Both authors
have greatly appreciated their interactions with the Harvard Philosophy
of Science group and its organizers Andrew Friedman and Elizabeth
Petrik, as well as with Scott Aaronson and Ned Hall. Both authors
are indebted to Steven Weinberg, who has generously exchanged relevant
ideas and whose own recent paper \cite{Weinberg:2014qmwsv} also explores
the idea of reformulating quantum theory in terms of density matrices.
The work of Pieter Vermaas has also been a continuing source of inspiration
for both authors, as have many conversations with Allan Blaer.
\end{acknowledgments}
\bibliographystyle{apsrev4-1}
\bibliography{Bibliography,BibliographyEntryPaper}

\end{document}